\documentclass[a4paper,11pt]{article}
\usepackage{pos}

\usepackage{afterpage,rotating}
\usepackage{currvita}

\usepackage{tabularx}
\usepackage{nonfloat}
\usepackage{graphicx}
\usepackage{dcolumn}
\usepackage{bm}
\usepackage{subfigure}
\usepackage{mathrsfs}
\usepackage{amsmath,amsbsy,lmodern}
\usepackage{epsfig,color}
\usepackage{placeins}
\usepackage{rccol}
\usepackage{comment}
\usepackage{hyperref} 
\usepackage{setspace} 
\usepackage{tabularx} 
\usepackage{multirow}
\usepackage{braket}
\usepackage{listings}
\usepackage{floatrow}
\usepackage{float}
\usepackage{accents}

\newcommand{\bs}{B_s^0}
\newcommand{\bsbs}{B_s^{*0} \bar{B}_s^{*0}$, $B_s^{*0} \bar{B}_s^0$ or $B_s^{0} \bar{B}_s^{*0}$, and $ B_s^0 \bar{B}_s^0}
\newcommand{\bsbsbara}{B_s^{(*)0}\bar{B}_s^{(*)0}}
\newcommand{\bseep}{B_s^0 \rightarrow \eta^\prime \eta}
\newcommand{\mbc}{M_{\rm bc} }
\newcommand{\de}{\Delta E}
\newcommand{\mep}{M(\pi^+\pi^-\eta)}
\newcommand{\bksep}{B^0 \rightarrow \eta^\prime K_S^0}

\title{Study of $B$ and $B_s$ Decays at Belle}

\author*[1]{N.~K.~Nisar}

\affiliation[]{Brookhaven National Laboratory,\\
  Upton, New York, 11973}
\note{On behalf of the Belle Collaboration}

\emailAdd{nnellikun@bnl.gov}

\abstract{
We report results on the search for $B_s\to\eta^{\prime}\eta$ decay, and the searches for $B^0\to\textrm{invisible}$ and $B^0\to\textrm{invisible}+\gamma$ decays.
 The former result is  based on a data sample of $121.4~\textrm{fb}^{-1}$ recorded at the $\Upsilon(5S)$ resonance while the later results are obtained from a $711~\textrm{fb}^{-1}$ of data sample collected at $\Upsilon(4S)$ resonance with the Belle detector at the KEKB $e^+e^-$ collider.
 We observe no significant signal for the decays and set upper limit on their branching fractions at 90\% confidence level of $\mathcal{B}(B_s\to\eta^{\prime}\eta)<7.1\times 10^{-5}$, $\mathcal{B}(B^0\to\textrm{invisible})<7.8\times 10^{-5}$ and $\mathcal{B}(B^0\to\textrm{invisible}+\gamma)<1.6\times 10^{-5}$.}

\FullConference{%
  40th International Conference on High Energy physics - ICHEP2020\\
  July 28 - August 6, 2020\\
  Prague, Czech Republic (virtual meeting)
}

\begin{document}
\maketitle

\section{Introduction}
In the Standard Model (SM), $B_s^0 \rightarrow \eta^\prime \eta$ decay
proceeds via tree-level $b\to u$ and penguin $b\to s$ transitions. 
Penguin transitions are sensitive to  Beyond-the-Standard-Model (BSM) physics scenarios
and could affect its branching fraction and {\it CP} asymmetry \cite{belleiiphysicsbook}.
Once the branching fractions for two-body decays $B_{s,d} \to \eta\eta, \eta\eta^{\prime}, \eta^{\prime}\eta^{\prime} $ are measured,
and the theoretical uncertainties are reduced,
it would be possible to extract {\it CP} violating parameters
from the data using the formalism based on SU(3)/U(3) symmetry~\cite{bf1}.
The formalism requires at least four of these six branching fractions and the result on $B_s^0 \rightarrow \eta^\prime \eta$ is a potential input.
The predicted branching fractions of the decays $B^0\to\textrm{invisible}$ and $B^0\to\textrm{invisible}+\gamma$, where “invisible” defined as particles that leave no signal in the
Belle detector, could be as high as $10^{-6}-10^{-7}$ in the New Physics (NP) models~\cite{ADedes,ABadin}. 
Decays with similar signature such as $B^0\to(\gamma)\nu\bar{\nu}$ and $B^0\to\nu\bar{\nu}\nu\bar{\nu}$ are highly suppressed in the SM~\cite{GBuchalla,BBhattacharya,CDLu}. A very low
background from the SM indicates that a signal of $B^0\to\textrm{invisible}+(\gamma)$ in the current B-factory data would indicate
NP.

\section{Belle detector}
The Belle detector~\cite{Belle}
was 
a large-solid-angle magnetic spectrometer
that operated at the KEKB asymmetric-energy $e^+e^-$ collider~\cite{KEKB}.
The detector components include
a tracking system comprising a silicon vertex detector (SVD) and a central drift chamber (CDC),
a particle identification (PID) system
that consists of a barrel-like arrangement of time-of-flight scintillation counters (TOF)
and an array of aerogel threshold Cherenkov counters (ACC),
and a CsI(Tl) crystal-based electromagnetic calorimeter (ECL).
All these components are located inside a superconducting
solenoid coil that provides a 1.5~T magnetic field.
Outside the coil, the
$K_L^0$ and muon detector (KLM) is
instrumented to detect $K_L^0$ mesons and to identify muons.

\section{Search for the Decay $B_s^0 \rightarrow \eta^\prime \eta$} 

In this paper we report the preliminary result of the first search for the decay $B_s^0 \rightarrow \eta^\prime \eta$
using the full Belle data sample of $121.4~\textrm{fb}^{-1}$
collected at the $\Upsilon(5S)$ resonance.
The $\Upsilon(5S)$ decays into $\bsbs$ pairs followed by the decays of
the excited vector states to $\bs$, by emitting a photon.
Our data sample contains $(6.53\pm0.66)\times10^{6}$ $\bsbsbara$ pairs~\cite{nbsbsb}.
A set of Monte Carlo (MC) simulated events are used for the selection optimization and estimation of reconstruction efficiency.

We reconstruct $\eta$ candidates
using pairs of photons of energy that exceeds 50 (100) MeV in the barrel (end-cap)
region of the ECL and requiring the invariant mass
to be in the range $515 \le M(\gamma\gamma) \le 580$~${\rm MeV/c}^2$.
  Candidates for the decay $\eta^{\prime}\to\pi^+\pi^-\eta$ are reconstructed
using pairs of oppositely-charged pions and $\eta$. We require the reconstructed $\eta^{\prime}$ 
invariant mass to be in the range $920\le M(\pi^+\pi^-\eta) \le 980$~${\rm MeV/c}^{2}$.
To identify $\bseep$ candidates we use
beam-energy constrained $B_s^0$ mass, $\mbc=\sqrt{E_{\rm beam}^2-p_{B_s}^2}$,
the energy difference, $\de=E_{B_s}-E_{\rm beam}$,
and the reconstructed invariant mass of the $\eta^\prime$,
where $E_{\rm beam}$, $p_{B_s}$ and $E_{B_s}$
are the beam energy,
the momentum magnitude
and the reconstructed energy of $B_s^0$ candidate,
respectively.

The primary source of background are $e^+e^-\to q\bar{q}$ ($q=u,d,c,s$) continuum events.
Because of large initial momenta of the light quarks,
continuum events exhibit a ``jet-like'' event shape,
while $\bsbsbara$ events are distributed isotropically.
We use modified Fox-Wolfram moments~\cite{ksfw}, which describe the topology of the event, 
to discriminate between signal and continuum background.

To extract the signal yield, we perform an unbinned extended maximum likelihood fit
to the three-dimensional (3D) distribution of $\mbc$, $\de$, and $\mep$.
MC sample is used to determine signal and background probability density functions (PDF).
 We use $B^0\to\eta^{\prime} K_S^0$ data recorded at the $\Upsilon(4S)$ resonance
to adjust the PDF shape parameters used to describe the signal.

To test and validate our fitting model, ensemble tests are performed by generating MC pseudoexperiments using PDFs obtained from the simulation and the $\bksep$ data.
We use the results of pseudoexperiments 
to construct classical confidence intervals 
using Neyman construction~\cite{frequentist_approach}. 
These confidence intervals are then used to prepare 
a classical confidence belt~\cite{belt_method} and used 
to make a statistical interpretation 
of the results obtained from fit to data.

We obtain $2.7 \pm 2.5$ signal and $57.3 \pm 7.8$ background events from the 3D fit to data.
We show the signal-region projections of the fit in Fig.~\ref{fit_data_Bseep}.
We observe no signal and estimate the 90\% confidence-level (CL) upper limit on the branching fraction
of the decay $\bseep$ using the frequentist approach~\cite{frequentist_approach} 
and the following formula:

\begin{figure}
\begin{center}
\includegraphics[width=1\textwidth]{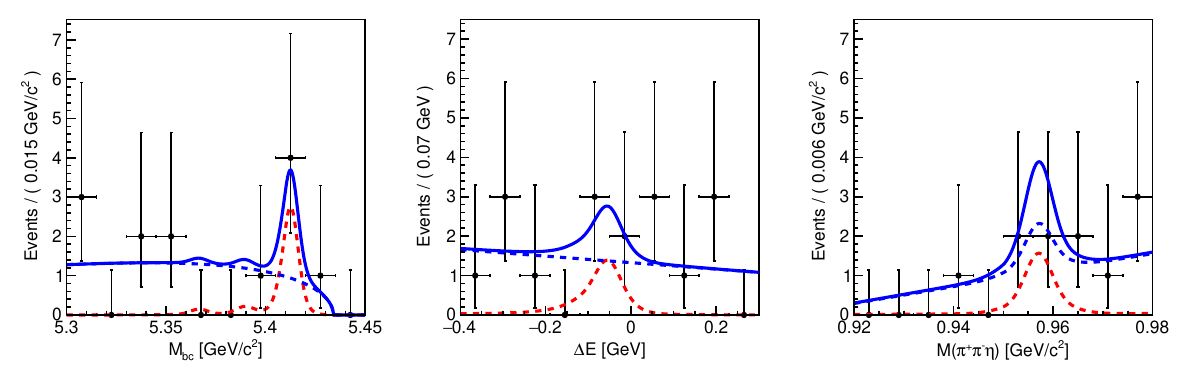}

\end{center}
\figcaption{Signal-region projections of 3D fit to $\bseep$ data.
Points with error bars represent data, blue
solid curves show the resulting fit-projection, while the red dash-dotted and blue dash-dotted curves show the signal and background components.}
\label{fit_data_Bseep}
\end{figure}

\begin{equation}
\mathcal{B}(\bseep) < \frac{N_{\textrm{UL}}^{90\%}}{2 \cdot N_{\bsbsbara} \cdot \varepsilon \cdot \mathcal{B}_{\textrm{dp}}},
\label{eq_ul}
\end{equation}

\noindent where $N_{\bsbsbara}$ is the number of $\bsbsbara$ pairs in the full Belle data sample,
$\varepsilon$ is the overall reconstruction efficiency for the signal $B_s^0$ decay,
and $\mathcal{B}_{\textrm{dp}}$ 
is the product of the secondary branching fractions for all daughter particles in our final state.
Further, $N_{\textrm{UL}}^{90\%}$ is the expected signal yield
at 90\% CL obtained from the confidence belt, which is approximately 6 events. Using Eq.~(\ref{eq_ul}) we
obtain a 90\% CL upper limit on the branching fraction
of $\mathcal{B}(\bseep) < 7.1 \times 10^{-5}$.
The total systematic uncertainty on the upper limit is estimated to be 17\%.

\section{Search for $B^0$ decays to invisible final states $(+\gamma)$} 

These searches are based on a data sample containing 
$772\times10^6$ $B\bar{B}$ pairs accumulated at the $\Upsilon(4S)$ resonance, 
corresponding to an integrated luminosity of
$711~\textrm{fb}^{-1}$.  Ten million MC simulated events 
for $B^0\to\nu\bar{\nu}$ and $B^0\to\nu\bar{\nu}\gamma$ decays are generated and 
used to determine signal efficiency and optimize signal event selection.

Since the signal side particle, except photon, cannot be detected, 
the other $B$ meson in the event ($B_{\textrm{tag}}$) is reconstructed.
Then the signal is searched in the remaining part of the event. 
$B_{\textrm{tag}}$ mesons are reconstructed from 494 hadronic decay modes by assigning signal probability to reconstructed particles using a
neural network (NN) package~\cite{full_rec}. 
After reconstruction of $B_{\textrm{tag}}$, 
no extra particles but photons are expected in the event. 
Thus events with extra tracks, $\pi^0$s, or $K_L^0$s are rejected.

The sum of all remaining energies of ECL clusters that are not 
associated with $B_{\textrm{tag}}$ daughters and signal photons in case of
 $B^0\to\textrm{invisible}+\gamma$, denoted by $E_{\textrm{ECL}}$,
 is a strong variable to identify signal events.
Since the distribution for signal events peaks at zero,
the $E_{\textrm{ECL}}$ signal box is defined as $E_{\textrm{ECL}}<0.3$ GeV
and the sideband is defined as $0.3 \textrm{GeV} < E_{\textrm{ECL}} < 1.2 \textrm{GeV}$.
Continuum events are the dominant source of background (Non-$B$) followed by $B\bar{B}$ decay
with a $b\to c$ transition (Generic $B$).
Two NNs are implemented to suppress these backgrounds.  

\begin{figure}
\begin{center}
\includegraphics[width=0.67\textwidth]{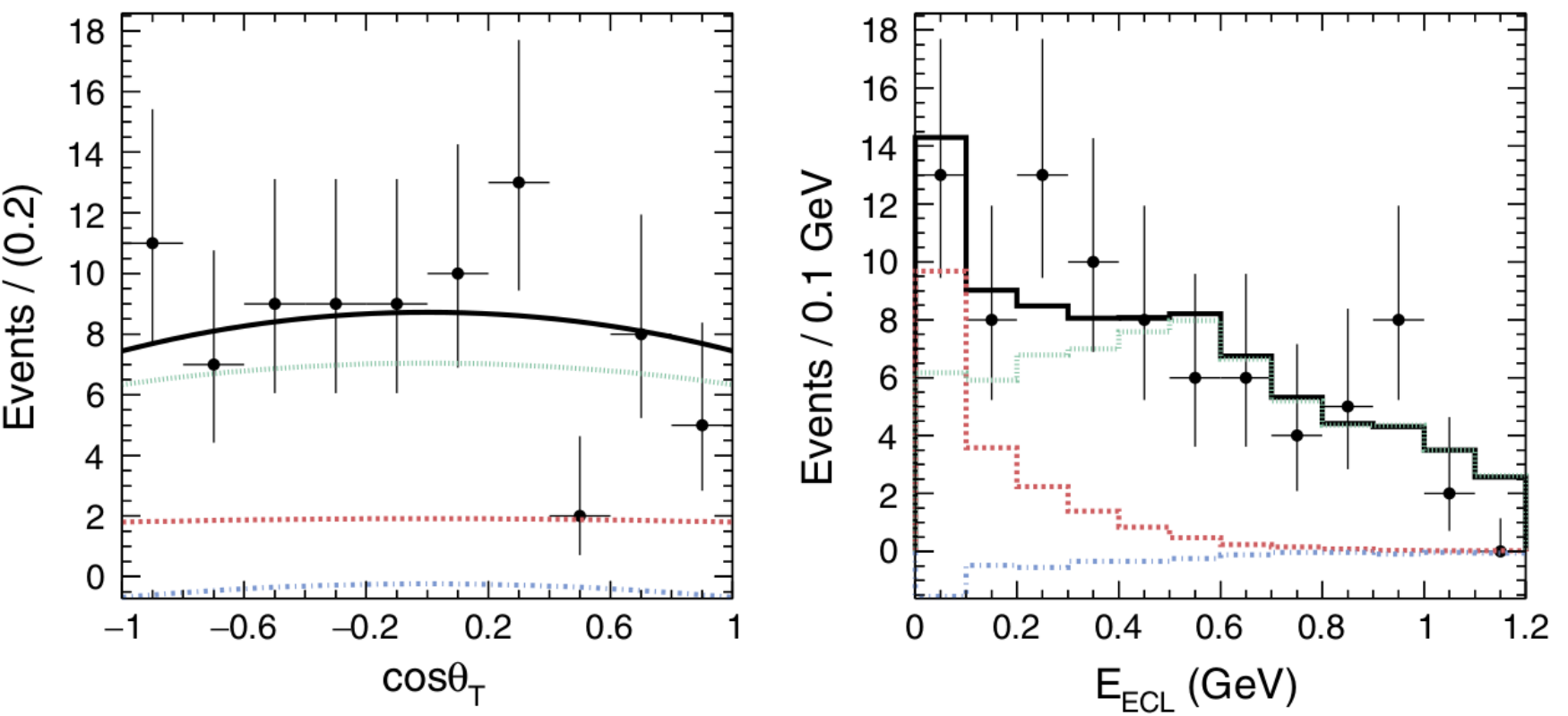}
\end{center}
\figcaption{Projections of the fit result on $\cos\theta_T$ (left) and $E_{\textrm{ECL}}$
(right) for $B^0\to\textrm{invisible}$. Points with error bars are data, black
solid line is the fit result, red dotted line is the signal component,
green short-dashed line is the generic $B$ background component
and blue dash-dotted line is the non-$B$ background component.}
\label{fit_data_Binv}
\end{figure}

A two dimensional (2D) extended unbinned maximum likelihood fit is
 applied with fitting variables $E_{\textrm{ECL}}$ and $\cos\theta_T$ to extract signal
 yield for the decay $B^0\to\textrm{invisible}$.
 Here $\cos\theta_T$ is the cosine of the angle between the two thrust axes in the $e^+e^-$ c.m. frame.
 The two thrust axes are defined as the directions
 that maximizes the longitudinal momenta of $B_{\textrm{tag}}$ daughters
 and particles in the remaining part of the event.
 All PDFs are obtained from signal MC and off-resonance data.
The projections of the 2D fit results are shown in Fig.~\ref{fit_data_Binv} and the corresponding fitting yiels for each component 
are listed in Table.~\ref{tab:BtoInv}. 
No significant signal is observed and a 90\% CL upper limit on the branching fraction is estimated to be $\mathcal{B}(B^0\to\textrm{invisible})<7.8\times 10^{-5}$~\cite{BtoInvGamma}.
Systematic uncertainty is estimated to be $7.9\%$ using control samples $B^{0,\pm}\to B^{*,\pm}l\nu$.

\begin{table}
\centering
\begin{tabular}{lc}
\hline \hline
Component     & Yields  \\
\hline 

Signal        & $18.8^{+15.3}_{-14.5}$ \\
Generic $B$   & $68.1^{+12.2}_{-11.7}$  \\
Non-$B$       & $-3.9^{+19.5}_{-17.5}$  \\
\hline
\hline
\end{tabular}
\caption{Fitting yield ($B^0\to\textrm{invisible}$).}
\label{tab:BtoInv}
\end{table}

$B^0\to{\textrm{invisible}}+\gamma$ decays are searched by counting events in $E_{\textrm{ECL}}$ signal box
 in the bins of squared missing mass defined as:

\begin{equation}
M_{\textrm{miss}}^2=(\vec{P}_{\textrm{beam}}- \vec{P}_{B_{\textrm{tag}}}-\vec{P}_{\gamma})^2/c^2,
\end{equation}
where $\vec{P}_{\textrm{beam}}$, $\vec{P}_{B_{\textrm{tag}}}$ and
$\vec{P}_{\gamma}$ are four-momenta of $e^+e^-$ system, the $B_{\textrm{tag}}$ and the signal photon. 
The number of background events in the $E_{\textrm{ECL}}$ signal box 
is estimated from the data sideband 
by multiplying the fraction
of events in signal box to the sideband, estimated in the MC.
The counting results in $E_{\textrm{ECL}}$ signal box and in bins of $M_{\textrm{miss}}^2$
are summarized in Table.~\ref{tab:BtoInvg}.  The observed number of events is consistant with no signal.  
We set a 90\% CL upperlimit on the branching fraction $\mathcal{B}(B^0\to\textrm{invisible}+\gamma)<1.6\times 10^{-5}$~\cite{BtoInvGamma} with an associated systematic uncertainty of $8.4\%$.


\begin{table}
\centering
\begin{tabular}{lccc}
\hline \hline
    & $N^{\textrm{data}}_{\textrm{bkg,box}}$  & $N^{\textrm{data}}_{\textrm{box}}$ \\
\cline{2-3}
 $B^0\to{\textrm{invisible}}+\gamma$ & $16.1\pm6.3$ & 11 \\
\hline
$M_{\textrm{miss}}^2<5~{\rm GeV}^2/c^4$  & $3.2\pm2.1$&2 \\
$5~{\rm GeV}^2/c^4<M_{\textrm{miss}}^2<10~{\rm GeV}^2/c^4$  & $1.0\pm0.8$&2 \\
$10~{\rm GeV}^2/c^4<M_{\textrm{miss}}^2<15~{\rm GeV}^2/c^4$   & $4.4\pm2.6$&3 \\
$15~{\rm GeV}^2/c^4<M_{\textrm{miss}}^2<20~{\rm GeV}^2/c^4$   & $7.1\pm2.9$&4 \\
$20~{\rm GeV}^2/c^4<M_{\textrm{miss}}^2$  & $6.6\pm2.9$&7 \\
\hline
\hline
\end{tabular}
\caption{Estimated number of background events in the
signal box ($N^{\textrm{data}}_{\textrm{bkg,box}}$) and the number of events in the signal box ($N^{\textrm{data}}_{\textrm{box}}$\
) for $B^0\to{\textrm{invisible}}+\gamma$ and $M_{\textrm{miss}}^2$ bins.}
\label{tab:BtoInvg}
\end{table}

\section{Conclusions}
In  summary,  we  have  used  the  full  data  sample  recorded  by  the  Belle  experiment
 at $\Upsilon(5S)$ and $\Upsilon(4S)$  resonances  to  search  for  the  decays
 $B_s^0 \rightarrow \eta^\prime \eta$ and $B^0\to\textrm{invisible}+(\gamma)$ and no evidence is found.
We set world’s first upper limits on the branching fraction of $B_s^0 \rightarrow \eta^\prime \eta$ and improved the existing upper limit on $B^0\to\textrm{invisible}+\gamma$.


\begin{thebibliography}{99}

\bibitem{belleiiphysicsbook}
E.~Kou {\it et al.} 
(Belle~II Collaboration), 
Prog Theor Exp Phys {\bf 2019}, 123C01 (2019). 

\bibitem{bf1}
Y.-K.~Hsiao, C.-F.~Chang, and X.-G.~He,
Phys. Rev. D {\bf 93}, 114002 (2016). 

\bibitem{ADedes}
A. Dedes, H. Dreiner, and P. Richardson, Phys. Rev. D {\bf65},
015001 (2001).
\bibitem{ABadin}
A. Badin and A. A. Petrov, Phys. Rev. D {\bf82}, 034005 (2010).

\bibitem{GBuchalla}
G. Buchalla and A. J. Buras, Nucl. Phys. {\bf B400}, 225 (1993).

\bibitem{BBhattacharya}
B. Bhattacharya, C. M. Grant, and A. A. Petrov, Phys. Rev. D {\bf99}, 093010 (2019).

\bibitem{CDLu}
C. D. Lu and D. X. Zhang, Phys. Lett. B {\bf381}, 348 (1996).

\bibitem{Belle}
A.~Abashian {\it et al.} (Belle Collaboration), Nucl. Instrum. Methods 
Phys. Res. Sect. A {\bf 479}, 117 (2002); also see Section 2 in
J.~Brodzicka {\it et al.}, Prog. Theor. Exp. Phys. {\bf 2012}, 04D001 (2012).

\bibitem{KEKB}
S.~Kurokawa and E.~Kikutani, Nucl. Instrum. Methods Phys. Res. Sect. A {\bf 499}, 1 (2003),
and other papers included in this Volume;
T.~Abe {\it et al.}, Prog. Theor. Exp. Phys. {\bf 2013}, 03A001 (2013) and references therein.

\bibitem{nbsbsb}
C. Oswald {\it et al.} (Belle Collaboration), 
Phys. Rev. D {\bf 92}, 072013 (2015). 


\bibitem{ksfw}
The Fox-Wolfram moments were introduced in G.~C.~Fox 
and S.~Wolfram, Phys. Rev. Lett. {\bf 41}, 1581 (1978). The
Fisher discriminant used by Belle, based on modified Fox-Wolfram moments, 
is described in K.~Abe {\it et al.}
(Belle Collaboration), 
Phys. Rev. Lett. {\bf 87}, 101801 (2001)
and 
K.~Abe {\it et al.} (Belle Collabboration.), Phys. Lett. B {\bf 511}, 151 (2001). 

\bibitem{frequentist_approach}
J.~Neyman, Phil. Trans. Roy. Soc. Lond. {\bf A236}, 767, 333 (1937); 
Reprinted in 
{\it A Selection of Early Statistical Papers of J. Neyman}, 
(University of California Press, Berkeley, 1967).

\bibitem{belt_method}
A.~Stuart and J.K.~Ord, {\it Classical Inference and Relationship},
5th ed., 
Kendall's Advanced Theory of Statistics, 
Vol.~2 (Oxford University Press, New York, 1991); 
see also earlier editions by Kendall and Stuart. \\
W.T.~Eadie, D.~Drijard, F.E.~James, M.~Roos, and B.~Sadoulet, 
{\it Statistical Methods in Experimental Physics}, 
(NorthHolland, Amsterdam, 1971). 
\bibitem{full_rec}
M.~Feindt, F.~Keller, M.~Kreps, T.~Kuhr, S.~Neubauer, D.~Zander, and A.~Zupanc, 
Nucl. Instrum. Methods Phys. Res., Sect. A {\bf654}, 432 (2011).

\bibitem{BtoInvGamma}
Y. Ku {\it et al.} (Belle Collaboration),
Phys. Rev. D {\bf 102}, 012003 (2020).








\end{thebibliography}
\end{document}